\documentclass[aps,preprint]{revtex4}%
\usepackage{amsfonts}
\usepackage{amsmath}
\usepackage{amssymb}
\usepackage{graphicx}%
\providecommand{\U}[1]{\protect\rule{.1in}{.1in}}

\begin{document}
\title{Critical temperature and specific heat for Cooper pairing on a spherical surface}
\author{V. N. Gladilin$^{1,\ast}$}
\author{J. Tempere$^{1,2}$}
\author{I. F.\ Silvera$^{2}$}
\author{J. T. Devreese$^{1}$}
\affiliation{$^{1}$ TFVS, Universiteit Antwerpen, Universiteitsplein 1, B-2610 Antwerpen, Belgium}
\affiliation{$^{2}$ Lyman Laboratory of Physics, Harvard University, Cambridge MA 02138, USA}
\pacs{PACS number}

\begin{abstract}
Based on an exact solution of the Bardeen-Cooper-Schrieffer type Hamiltonian
on a spherical surface, we calculate the specific heat for the electron system
with pair correlations on a sphere. We find that it is possible to extract
from the specific heat a temperature above which many-body states with broken
Cooper pairs get populated. Therefore, we define this temperature as the
characteristic temperature signalling the onset of a BCS-type pair-correlated
state for electrons on a spherical surface. Such spherical electron systems
are realized in multielectron bubbles in liquid helium, for which the
above-mentioned characteristic temperature is found to be of the order of
10-100 mK. Both the specific heat and the critical temperature show a
pronounced (4-6\%) odd-even parity effect that persists even for numbers of
electrons as large as 10$^{6}$.

\end{abstract}
\date[Date text]{date}
\maketitle

\section{Introduction}

The physics of the two-dimensional electron system remains the subject of
intense theoretical and experimental scrutiny. Spherical two-dimensional
electron systems have not been as thoroughly investigated as their flat
counterparts, despite the fact that the spherical geometry is expected to lead
to new physics because of its intrinsic topological difference with a flat
space\cite{HairyBall}. Moreover, the presence of curvature also influences the
coupling of the system to magnetic fields, as has been shown for
spherical\cite{GokerJPHYS16} and toroidal electron
systems.\cite{EncinosaPRA73}

Spherical two-dimensional electron systems appear in various physical systems:
prominent examples are metallic nanoshells \cite{Nanoshells} and
buckyballs\cite{AmovilliPRA66}. The most idealized realization of a spherical
two-dimensional electron system is found in a multielectron bubble (MEB) in
liquid helium. MEBs are (typically micron-sized) cavities in the liquid,
containing anywhere from a few up to 10$^{8}$ electrons. They have first been
observed as a result of a surface instability that occurs when a liquid helium
surface is being charged with electrons beyond a critical surface charge
density.\cite{MEB}

{The bare single electron states on the surface of a spherical bubble are
angular momentum eigenstates and have discrete energies, characterized by the
angular momentum $L$ and with degeneracy $2L+1$. At low temperature there will
be a well defined Fermi surface located at the highest occupied state.
Small-amplitude shape oscillations, including surface waves, can be quantized
as spherical ripplons, described by the Hamiltonian
\begin{equation}
\hat{H}_{\text{ripl}}=\sum_{L>0,m}\hbar\omega_{L}\hat{a}_{L,m}^{+}\hat
{a}_{L,m}, \label{Hrip}%
\end{equation}
where $\hat{a}_{L,m}^{+}$ ($\hat{a}_{L,m}$) is the creation (annihilation)
operator for a ripplon with angular momentum $L$ and its $z$-projection $m$.
$\omega_{L}$ are the ripplonic frequencies derived in~\cite{TemperePRL87} for
spherical MEBs under external pressure. The electrons can interact with these
ripplons.} The interaction between electrons and ripplons on the bubble
surface is analogous to that of electrons and phonons in solids. In
particular, it can lead to the formation of ripplonic
polarons.\cite{Riplopols} In a recent work, the present authors have shown
that the electron-ripplon interaction leads to the formation of strong pairing
correlations below a critical temperature.\cite{TemperePRB72} These pairing
correlations are similar to the Bardeen-Cooper-Schrieffer (BCS)
superconducting pair correlations in metals. This raises the obvious question
of how \textquotedblleft superconductivity\textquotedblright\ will manifest
itself on a micron-sized spherical surface. More specifically, one can ask how
the formation of strong pairing correlations can be detected experimentally,
and how the critical temperature can be determined.

In their study of nanoscopic superconducting aluminium particles, Black
\emph{et al}. \cite{BlackPRL76} used electron tunneling spectroscopy to detect
the superconducting excitation gap. Also in multielectron bubbles, pairing
correlations lead to a (pseudo)gap in the density of
states.\cite{TemperePRB72} {In the present paper, we investigate observable
manifestations of pairing correlations in MEBs. The altered density of states
will reflect itself in particular in the specific heat of the system. Below,
we calculate the specific heat of the spherical two-dimensional electron
system and argue that the onset of a sharp rise in the specific heat can be
used to define a characteristic temperature for the transition between a
normal state and the pair-correlated state. Moreover, we derive an analytical
expression for this characteristic temperature and show that the specific heat
reveals a pronounced odd-even parity effect, which persists even for large
($10^{6}$) numbers of electrons.}

\section{Specific heat of Cooper pairs on a sphere}

The spherical, two dimensional electron system, is described in second
quantization with the operators $\hat{c}_{\ell,m,\sigma}^{+}$ and $\hat
{c}_{\ell,m,\sigma}$ that create resp. destroy an electron with spin
$\sigma=\uparrow,\downarrow$ in the angular momentum eigenstate $\left\vert
\ell,m\right\rangle $. In Ref. \cite{TemperePRB72} we have shown that in the
case of multielectron bubbles, the electron-ripplon interaction leads to a
Cooper-type attractive interaction between the electrons, resulting in a
BCS-like Hamiltonian:
\begin{align}
\hat{H}  &  =%
{\displaystyle\sum\limits_{\ell=0}^{\infty}}
\left(  \sum_{m=-\ell}^{\ell}\sum_{\sigma}\epsilon_{\ell}\hat{c}%
_{\ell,m,\sigma}^{+}\hat{c}_{\ell,m,\sigma}\right. \nonumber\\
&  \left.  -G\sum_{m,m^{\prime}=-\ell}^{\ell}\hat{c}_{\ell,-m^{\prime
},\downarrow}^{+}\hat{c}_{\ell,m^{\prime},\uparrow}^{+}\hat{c}_{\ell
,m,\uparrow}\hat{c}_{\ell,-m,\downarrow}\right)  . \label{ham}%
\end{align}
This Hamiltonian can be mapped onto the so-called "reduced BCS Hamiltonian"
studied in the context of superconducting
nanograins.\cite{vonDelftPRL77,MatveevPRL79} In the present case,
$\epsilon_{\ell}$ represents the unperturbed, $2(2\ell+1)$-degenerate
angular-momentum energy level
\begin{equation}
\epsilon_{\ell}=\frac{\hbar^{2}}{2m_{e}R^{2}}\ell(\ell+1),
\end{equation}
where $m_{e}$ is the electron mass and $R$ is the radius of the spherical
system. The separation between different angular momentum levels $\ell$ of the
unperturbed system is typically much larger than the energy of the ripplons,
and the interacting electron-ripplon system can effectively be treated as a
sum over subsystems with different $\ell$. The typical energy scale of the
kinetic energy is $\epsilon_{1}=\hbar^{2}/(m_{e}R^{2})$. For a 10000 electron
bubble at zero external pressure, $\epsilon_{1}=0.78$ mK. The interaction
energy scale $G$ for the same bubble lies around $G=10$ mK, and the ripplon
energy scale for this bubble is $10$ $\mu$K. In what follows, we will present
results as a function of the dimensionless coupling constant $G/\epsilon_{1}$.

The Hamiltonian (\ref{ham}) can be solved analytically, using Richardson's
method.\cite{RichardsonPL3,RichardsonNP52} The total energy is a sum of the
energies for each $\ell$ subsystem,
\begin{equation}
E_{j}=%
{\displaystyle\sum\limits_{\ell=0}^{\infty}}
E_{n_{\ell},g_{\ell},b_{\ell}}^{(\ell)},
\end{equation}
and is characterized by the set of quantum numbers $j=\{n_{\ell},g_{\ell
},b_{\ell}\}_{\ell=0,1,2,...}.$ The energy levels of the subsystem $\ell$ are
given by
\begin{equation}
E_{n_{\ell},g_{\ell},b_{\ell}}^{(\ell)}=(2n_{\ell}+b_{\ell})\epsilon_{\ell
}-G(n_{\ell}-g_{\ell})(2\ell-b_{\ell}+2-n_{\ell}-g_{\ell}).\label{energies}%
\end{equation}
The quantum number $n_{\ell}$ corresponds to the number of electron pairs in
the subsystem, whereas $b_{\ell}$ is the number of unpaired electrons and
$g_{\ell}$ is the number of elementary bosonic pair-hole
excitations\cite{RomanPRB67} in the system of $n_{\ell}$ pairs. In the ground
state all electrons are paired ($b_{\ell}=0$), except one when an odd number
of electrons is present in the system. There are no pair-hole excitations
($g_{\ell}=0$) in the ground state. The spherical symmetry, resulting in a
discrete and degenerate single-particle level structure, allows for an exact
solution of the Hamiltonian (\ref{ham}), describing also the fluctuations
\cite{DiLorenzoPRL84}. At finite temperatures, two types of excitations will
be present. The broken pair states ($b_{\ell}\neq0$) correspond on a
mean-field level to quasiparticle states in the BCS approximation.\ The
presence of collective modes (in the context of the Richardson solution
sometimes called `gaudinos') correspond to the case when $g_{\ell}\neq0$
\cite{DukelskyRMP76}$.$

At strong coupling ($G\geqslant\epsilon_{1}/2$) the electron pairs
redistribute themselves over approximately $\mu=2G/\epsilon_{1}$ different
$\ell$ subsystems around the Fermi level $\ell=L_{F}$ to minimize the
energy.\cite{TemperePRB72} {. The above estimate for the number of partially
occupied angular momentum levels $\ell$ does not hold for $G\gtrsim
N\epsilon_{1}/2$, where $N$ is the total number of electrons in the system. As
implied from Eq.~(25) of Ref.~\onlinecite{ TemperePRB72}, at $G\gtrsim
N\epsilon_{1}/2$, when the interaction strength $G$ exceeds the Fermi energy
for non-interacting electrons, the ground state would correspond to an
electron configuration with just a single pair per $\ell$-level. However, such
an extreme strong-coupling limit appears irrelevant for the multielectron
bubbles under consideration and it is not addressed in the following, where
the inequality $G\ll N\epsilon_{1}/2$ is assumed to be satisfied.}

The degeneracy of energy level $E_{j}$ is a product of the degeneracies of the
energy levels of the constituent subsystems:%
\begin{equation}
J_{j}=%
{\displaystyle\prod\limits_{\ell=0}^{\infty}}
J_{n_{\ell},g_{\ell},b_{\ell}}^{(\ell)},
\end{equation}
with
\begin{equation}
J_{n_{\ell},g_{\ell},b_{\ell}}^{(\ell)}=2^{b_{\ell}}C_{b_{\ell}}^{2\ell
+1}\times\left\{
\begin{array}
[c]{lc}%
1\text{ } & g_{\ell}=0\\
\left(  C_{g_{\ell}}^{2\ell+1-b_{\ell}}-C_{g_{\ell}-1}^{2\ell+1-b_{\ell}%
}\right)  & g_{\ell}\geqslant1
\end{array}
\right.  \label{degen}%
\end{equation}
where $C_{n}^{k}$ are binomial coefficients.\cite{GladilinPRB70}

\begin{figure}
[ptb]
\begin{center}
\includegraphics[
height=3.2453in,
width=5.3134in
]%
{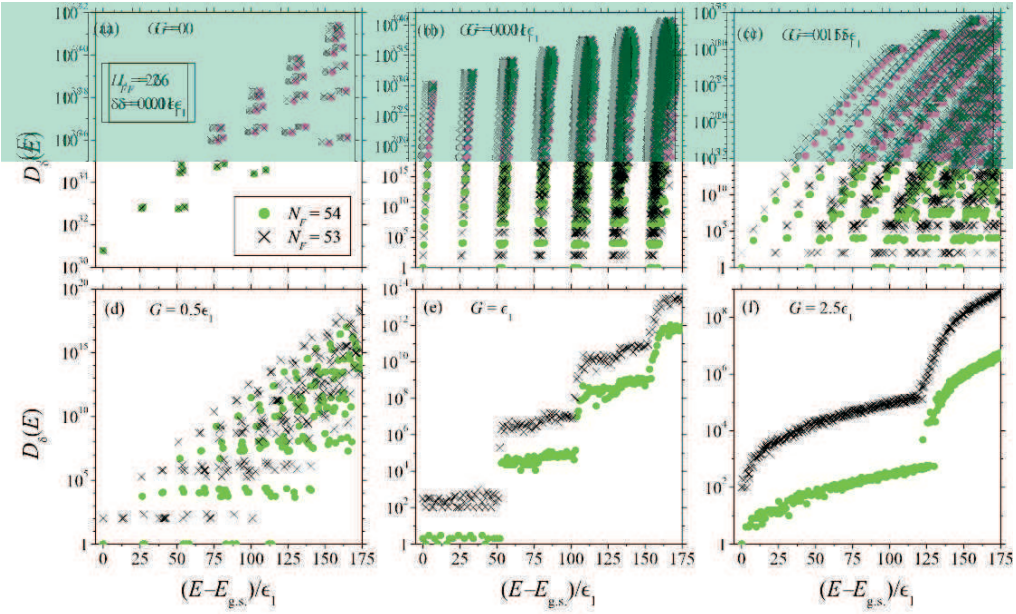}%
\caption{The calculated density of states for a spherical 2D electron system
with pairing correlations is shown as a function of the energy above the
ground state energy $E_{gs}$. Here $D_{\delta}(E)$ is evaluated by counting
the number of states in an interval $\delta=0.01\epsilon_{1}$ around energy
$E$. Results are shown for even ($N_{F}=54$, solid circles) and odd
($N_{F}=53$, crosses) numbers of electrons on the (unperturbed) Fermi angular
momentum level $L_{F}=26$. The different panels correspond to different
strengths of the attractive electron-electron term in the BCS-type Hamiltonian
(\ref{ham}).}%
\label{fig1}%
\end{center}
\end{figure}

The analytic expression for the energy levels and their degeneracies allow to
compute the specific heat straightforwardly%
\begin{equation}
C=\frac{d\left\langle E\right\rangle }{dT},
\end{equation}
where $\left\langle E\right\rangle $ is the statistical average energy of the
electrons in the MEB:%
\begin{equation}
\left\langle E\right\rangle =\frac{\sum_{j}J_{j}E_{j}\exp\left[
-E_{j}/\left(  k_{B}T\right)  \right]  }{\sum_{j}J_{j}\exp\left[
-E_{j}/\left(  k_{B}T\right)  \right]  }.
\end{equation}
The density of states $D_{\delta}(E)$ allows to express this as an energy
integral. The subscript $\delta$ is used to indicate that we numerically
calculate the density of states by counting the states in an energy interval
$\delta$ around the energy $E$. For the results presented in Fig. \ref{fig1},
we chose energy intervals of $\delta=0.01$ $\epsilon_{1}$. Fig. \ref{fig1}
shows the density of states $D_{\delta}(E)$ as a function of the energy above
the ground state energy $E_{gs}$, for different values of the coupling
constant $G/\epsilon_{1}$. Each graph shows results for both an even (full
circles) and an odd (crosses) number $N_{F}$ of electrons on the highest
angular momentum level $L_{F}$ occupied at $G=0$. As the coupling constant is
increased, a clear odd-even effect appears in the density of states. The
presence of an unpaired electron sensitively raises the degeneracy of the
energy levels and thus the number of available states.%

The aim of the next section is to examine the effect of pairing interactions,
described by the Hamiltonian (\ref{ham}) and characterized by the coupling
strength $G$, on the behavior of the specific heat as a function of temperature.

\bigskip

\section{Results and Discussion}

\subsection{Small coupling, $G/\epsilon_{1}<1/2$}

\begin{figure}
[ptb]
\begin{center}
\includegraphics[
height=2.2897in,
width=2.4168in
]%
{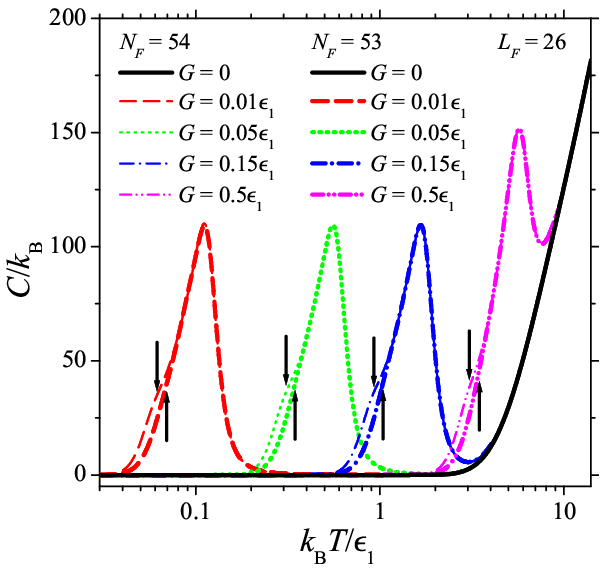}%
\caption{The calculated specific heat is plotted as a function of temperature
for an even and odd number of electrons. The results are shown for several
values of the interaction strength $G\leqslant\epsilon_{1}/2$. The down and up
arrows are positioned at the corresponding characteristic temperatures given
by (\ref{Teven})\ and (\ref{Todd}) for even and odd MEBs, respectively.}%
\label{fig2}%
\end{center}
\end{figure}

When there are no pairing interactions ($G=0$), the energy level spectrum
consists of the discrete levels $\epsilon_{\ell}$ and the specific heat will
be similar to that of a collection of fermionic quantum rotors. It will be
small for temperatures $T\ll\hbar^{2}/(2mR^{2}k_{B})$ and rise rather abruptly
to saturate at $Nk_{B}$ for $T\gg\hbar^{2}/(2mR^{2}k_{B})$. The onset of the
increase of the specific heat at $G=0$ is shown in the full curve in Fig.
\ref{fig2}. The increase in the specific heat starts at a temperature where
there appears a non-negligible occupation probability for the lowest excited
states. These excited states are separated from the ground state by an energy
$\sim$ $\epsilon_{1}(L_{F}+1)$ and correspond to transitions of electrons
between different $\ell$ subsystems. These transitions will be referred to as
\emph{interlevel} transitions.%

For small, non-zero $G\ll\epsilon_{1}$, an additional peak in $C(T)$ develops
as shown in Fig. \ref{fig2}. This additional peak appears at a temperature
smaller than the temperature at which the $G=0$ specific heat starts to
increase. This peak corresponds to \emph{intralevel} excitations of the
system: pair breaking and pair-hole excitations. As follows from Eq.
(\ref{energies}), the first pair-breaking energy equals the lowest pair-hole
excitation energy. For a subsystem $\ell$ with an even number of electrons, we
find $\Delta_{\ell}=(2\ell+1)G$. For an odd number of electrons, this becomes
$\Delta_{\ell}=2\ell G$.

So, when $G\ll\epsilon_{1}$, the intralevel excitations occur at energies much
smaller than the interlevel excitations. As $G$ is increased, the specific
heat peak that corresponds to these intralevel excitations shifts to higher
temperatures. In the case of a closed-shell configuration, no intralevel
excitations from the ground state are possible, so that no additional peak of
$C(T)$ appears at $G\ll\epsilon_{1}$.

\subsection{Large coupling, $G/\epsilon_{1}>1/2$}

\begin{figure}
[ptb]
\begin{center}
\includegraphics[
height=2.314in,
width=2.3997in
]%
{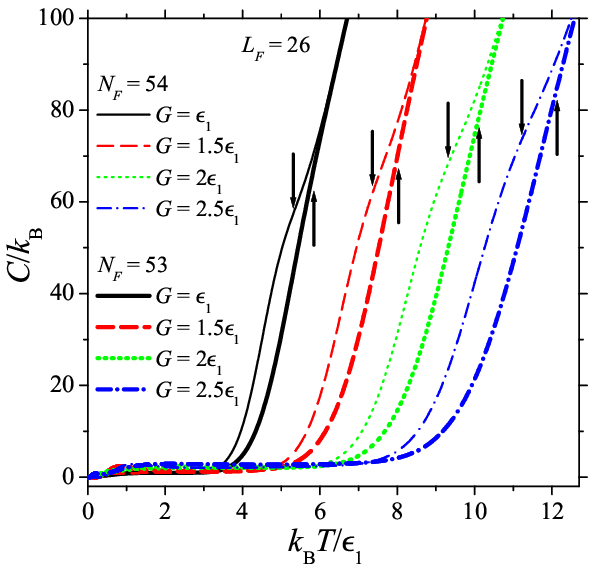}%
\caption{The calculated specific heat is plotted as a function of temperature
for an even and odd number of electrons. The results are shown for several
values of the interaction strength $G>\varepsilon_{1}/2$. The down and up
arrows are positioned at the corresponding characteristic temperatures given
by (\ref{Teven}) and (\ref{Todd}) for even and odd MEBs, respectively.}%
\label{fig3}%
\end{center}
\end{figure}

For $G>\epsilon_{1}/2$, the pair-breaking energy becomes larger than the
energy spacing between the single-electron $\ell$ levels. The density of
states no longer resembles a set of peaks, but it becomes more similar to a
step function, where the steps occur at the pair-breaking energy. When the
temperature is small enough so as not to populate states with a broken pair,
the available density of states is small. But when the temperature is
increased and broken pair states become populated, the relevant density of
states jumps to a value that is orders of magnitude higher (cf. Figs.
\ref{fig1}e, \ref{fig1}f). This reflects itself in the behavior of the
specific heat: as soon as the temperature is large enough to break pairs, the
specific heat will be strongly enhanced. This abrupt transition can be seen in
Fig. \ref{fig3}.%

The values of the temperature that correspond to the fast increase in $C(T)$
clearly correlate with $G$. {Note that -- as distinct from the case of
$G/\epsilon_{1}<1/2$ -- at large coupling ($G/\epsilon_{1}>1/2$) the specific
heat is non-zero even at temperatures below the temperature of the onset of
the rise in $C$. This is due to the presence of excitations related to
interlevel transitions of \emph{pairs}: for large $G$, these can have energies
smaller than the pair-breaking energy. With increasing $G$, the number of
these excitations increases (cp. Fig.~\ref{fig1}f to Fig.~\ref{fig1}e),
resulting in the corresponding increase of $C$ with $G$ in the low-temperature
region (see Fig.~3)}.

\subsection{The critical temperature for pair-breaking}

As implied form the previous discussion and from Figs. \ref{fig2}, \ref{fig3},
both for the case of small ($G<\epsilon_{1}/2$) and large ($G\geqslant
\epsilon_{1}/2$) coupling the specific heat starts rising at a temperature
when states with broken pairs get populated. Below this temperature, BCS-type
pairing correlations are dominant, but above this temperature, the
"superconducting" pairing correlations are suppressed. So, it is possible to
extract from $C(T)$ a typical temperature $T^{\ast}$ separating the two
aforementioned regimes.

Here we make an analytical estimate for $T^{\ast}$. The probability for the
system to be in the ground state is proportional to $P_{gs}=J_{gs}\exp\left(
-E_{gs}/k_{B}T\right)  $, where $J_{gs}$ is the degeneracy of the ground
state. The probability for the system to break up one pair is proportional to
$P_{1bp}=J_{1bp}\exp\left[  -\left(  E_{gs}+\Delta_{L_{F}}\right)
/k_{B}T\right]  $, since $\Delta_{L_{F}}$ is the first pair-breaking energy.
Here, $J_{1bp}$ is the total number of states with one broken pair. An
appreciable contribution of states with one broken pair to the specific heat
appears when $P_{1bp}$ becomes comparable to $P_{gs}$, i.e. at a temperature%
\begin{equation}
k_{B}T^{\ast}=\frac{\Delta_{L_{F}}}{\ln(J_{1bp}/J_{gs})}. \label{tstar1}%
\end{equation}
The ground state for an even bubble is characterized by $g_{\ell}=0$,
$b_{\ell}=0.$ A state with one broken pair is characterized by $n_{\ell
}\rightarrow n_{\ell}-1$ and $b_{\ell}\rightarrow b_{\ell}+2.$ At small
coupling, the first pair-breaking takes place for $\ell=L_{F}$. At strong
coupling {($\epsilon_{1}/2<G\ll N\epsilon_{1}/2$)}, the electron pairs
partially occupy about $\mu=2G/\epsilon_{1}$ single-particle $\ell$ levels
around the Fermi level $L_{F}$ and the first pair-breaking can take place on
any of these levels (so we still have $\ell\approx L_{F}$ for $L_{F}\gg1$).

From (\ref{degen}) we see that the degeneracy of states that correspond to the
first intralevel pair-breaking exceeds that of the ground state by a factor
$J_{1bp}/J_{gs}=4C_{2}^{2L_{F}+1}$ ($\approx8L_{F}^{2}$ for $L_{F}\gg1$). The
ground state for an odd bubble is characterized by $g_{\ell}=0$, $b_{\ell}=1$
for the lowest partially occupied subsystem with $\ell\approx L_{F},$ so that
the odd bubble has a ground state that is a factor $2C_{1}^{2L_{F}+1}$ more
degenerate than that of the even bubble. Applying again Eq. (\ref{degen}) we
get for odd bubbles $J_{1bp}/J_{gs}=2C_{3}^{2L_{F}+1}/C_{1}^{2L_{F}+1}%
\approx8L_{F}^{2}/3$.

Given that the pair-breaking energy is equal to the pair-hole excitation
energy, why did we assume that the increase in the specific heat is due to the
breaking of pairs rather than to pair-hole excitations ? This is justified
because the increase in degeneracy due to the pair-breaking is much larger
than the increase in degeneracy due to a pair-hole excitation. A pair-hole
excitation is characterized by $g_{\ell}\rightarrow g_{\ell}+1$ (whereas
$b_{\ell}$ and $n_{\ell}$ are unchanged). From (\ref{degen}) we see that
indeed the increase in the degeneracy of states due to the lowest pair-hole
excitation (for an even bubble) is characterized by the factor $(C_{1}%
^{2L_{F}+1}-1)$, much smaller than $J_{1bp}/J_{gs}$ for $L_{F}\gg1$.

Thus far, we have only considered intralevel pair breaking. But for
$G>\epsilon_{1}/2$, also interlevel pair breaking, where an electron pair is
broken up and one electron is transferred to another $\ell$ subsystem, become important.

At strong coupling, the electron pairs partially occupy about $\mu
=2G/\epsilon_{1}$ single-particle $\ell$ levels around the Fermi level $L_{F}
$. All these $\mu$ single-particle levels become relevant in the pair-breaking
process, raising the degeneracy of states due to pair-breaking so that
$J_{1bp}/J_{gs}\approx4C_{2}^{\mu(2L_{F}+1)}\approx8L_{F}^{2}\mu^{2}$ for even
bubbles and similarly $J_{1bp}/J_{gs}\approx8L_{F}^{2}\mu^{2}/3$ for odd bubbles.

\subsection{Parity effect}

From the preceding discussion, it is clear that there appears a parity effect.
The presence of an unpaired electron in the odd bubbles alters the degeneracy
ratio of the broken-pair state over the ground state. The presence of the
unpaired electron also changes the pair-breaking energy to $\Delta_{L_{F}%
}=2L_{F}G$ as compared with $\Delta_{L_{F}}=(2L_{F}+1)G$ for the even bubble.
The estimate for the temperature $T^{\ast}$ at which the broken-pair states
start being populated will therefore also be different for even and odd cases.
In particular, we have for the even case
\begin{equation}
k_{B}T_{even}^{\ast}\approx\frac{(2L_{F}+1)G}{\ln(8L_{F}^{2}\mu^{2})},
\label{Teven}%
\end{equation}
with $\mu=\max[1,2G/\epsilon_{1}]$, and for the odd case%
\begin{equation}
k_{B}T_{odd}^{\ast}\approx\frac{2L_{F}G}{\ln(8L_{F}^{2}\mu^{2}/3)}.
\label{Todd}%
\end{equation}
The location of these temperatures is indicated by arrows in Figs. \ref{fig2},
\ref{fig3}. The estimates (\ref{Teven}),(\ref{Todd}) are in good agreement
with the temperatures at which $C(T)$ increases. Note that for even bubbles,
the rise of the specific heat seems to show an initial "shoulder", and
$T_{even}^{\ast}$ is a bit smaller than $T_{odd}^{\ast}$. This shoulder in the
specific heat corresponds to the first pair-breaking transition. What makes it
different from the following pair-breaking transitions? As implied from Fig.
\ref{fig1} depicting the density of states, the relative increase in
$D_{\delta}$ due to the first pair-breaking transition is larger than that for
further pair-breaking transitions. Moreover, the energy separation of states
with one broken pair from adjacent states is larger than that of states with
more than one broken pair.%

\begin{figure}
[ptb]
\begin{center}
\includegraphics[
height=5.2112in,
width=3.2274in
]%
{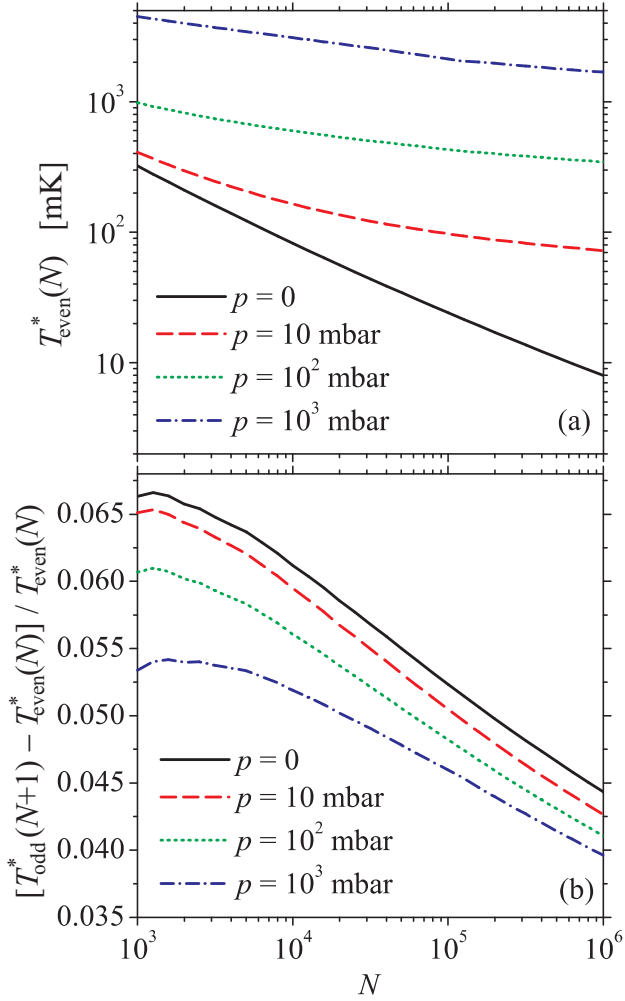}%
\caption{The characteristic temperature above which states with broken pairs
become populated is shown as a function of the number of electrons on the
sphere in panel (a) for an even number of electrons. In panel (b) the relative
difference between the characteristic temperatures of even and odd systems is
shown. Here $p$ is the pressure on a multielectron bubble. }%
\label{fig4}%
\end{center}
\end{figure}

{From Eqs.~(\ref{Teven}) and (\ref{Todd}), the relative difference between the
characteristic temperature for odd and even bubbles is
\begin{equation}
(T_{odd}^{\ast}-T_{even}^{\ast})/T_{even}^{\ast}=\left[  \frac{\ln(3)}%
{\ln(8L_{F}^{2}\mu^{2})}-\frac{1}{2L_{F}+1}\right]  \left[  1-\frac{\ln
(3)}{\ln(8L_{F}^{2}\mu^{2})}\right]  ^{-1}. \label{diff1}%
\end{equation}
For MEBs with large numbers of electrons ($N>>1$), the angular momentum
$L_{F}$ increases with $N$ approximately as $\sqrt{N/2}$. With increasing $N$,
also the parameter $\mu=2G/\epsilon_{1}$ rises (it is about 11.6 for
$N=10^{3}$ and 170 for $N=10^{6}$. Therefore, for sufficiently large $N$,
inequalities $\ln(3)\ll\ln(8L_{F}^{2}\mu^{2})\ll2L_{F}+1$ are satisfied so
that Eq.~(\ref{diff1}) takes the form
\begin{equation}
(T_{odd}^{\ast}-T_{even}^{\ast})/T_{even}^{\ast}\approx\ln(3)/\ln(8L_{F}%
^{2}\mu^{2}). \label{diff2}%
\end{equation}
The relative difference $(T_{odd}^{\ast}-T_{even}^{\ast})/T_{even}^{\ast}$,
described by Eq.~(\ref{diff2}), reflects the fact that -- as discussed in the
previous subsection -- for odd bubbles the ratio $J_{1bp}/J_{gs}$ is
approximately 3 times smaller as compared to that in even bubbles. With
increasing $N$, the ratio $J_{1bp}/J_{gs}$ strongly increases, while the
relative difference in $J_{1bp}/J_{gs}$ between odd and even grains remains
the same at any $L_{F}\gg1$. Since the characteristic temperature $T^{\ast}$
depends on $J_{1bp}/J_{gs}$ logarithmically [see Eq.~(\ref{tstar1})], the
effect of the aforementioned difference in $J_{1bp}/J_{gs}$ on $(T_{odd}%
^{\ast}-T_{even}^{\ast})/T_{even}$ weakens with increasing $N$. As a result,
the relative difference between the characteristic temperature for odd and
even bubbles decreases with increasing $N$, but slowly.} The characteristic
temperature for multielectron bubbles with an even number of electrons, is
shown in panel (a) of Fig. \ref{fig4} as a function of the number of electrons
in the bubble and the pressure exerted on the bubble. The pressure on a
multielectron bubble increases the electron-ripplon coupling, and thus $G$,
and $T^{\ast}$. The relative difference between the even and odd cases{, as
given by Eq.~(\ref{diff1}),} is shown in panel (b). Even for fairly large
numbers of electrons{, $N\sim10^{6}$, the $T_{odd}^{\ast}$ is 4\% higher than
$T_{even}^{\ast}$}.

\section{Conclusions}

In this paper, we have focused our discussion on the spherical electron system
as it is realized in multielectron bubbles, even though the results presented
here are more general, and valid for all the systems well described by the
Hamiltonian (\ref{ham}). This BCS-like Hamiltonian on the sphere can be solved
analytically using Richardson's method, and the specific heat is calculated
from the energy level spectrum and the level degeneracies, or from the density
of states. The ground state of the system is a state with strong BCS-like pair
correlations. We show that the specific heat shows a sharp increase as soon as
the temperature is large enough to populate the many-body states that contain
broken pairs. The temperature at which the specific heat starts to increase
constitutes a characteristic temperature $T^{\ast}$ that separates the
BCS-like pair-correlated state below $T^{\ast}$ from the broken-pair states
above $T^{\ast}$. This definition relates $T^{\ast}$ to the critical
temperature for the onset of superconductivity that can be derived from
measuring the specific heat of a superconducting sample. It is important to
note that the procedure followed here is not able to distinguish whether the
BCS-like pair-correlated state appears due to a phase transition or due to a
smooth evolution from the `normal' Fermi liquid. The presence of an unpaired
electron in spherical systems with an odd number of electrons significantly
affects both the specific heat and $T^{\ast}$: we find that for multielectron
bubbles with odd numbers of electrons $T^{\ast}$ is roughly 4-6\% larger than
in the even case. This persistent parity effect is present even for a large
total number of electrons.

\begin{acknowledgments}
The authors gratefully acknowledge stimulating discussions with V.M. Fomin.
J.T. is supported financially by the Fonds voor Wetenschappelijk Onderzoek -
Vlaanderen. This research has been supported financially by the FWO-V projects
Nos. G.0356.06, G.0115.06, G.0435.03, G.0306.00, the W.O.G. project
WO.025.99N, the GOA BOF UA 2000 UA. Part of this work was also supported by
the Department of Energy, Grant DE-FG02-ER45978. J.T. gratefully acknowledges
support of the Special Research Fund of the University of Antwerp, BOF\ NOI UA 2004.
\end{acknowledgments}

\bigskip

\bigskip

\end{document}